# Recent La Plata basin drought conditions observed by satellite gravimetry

J.L. Chen [1], C.R. Wilson [1,2], B.D. Tapley[1], L. Longuevergne [2,3], Z.L. Yang [2], B.R. Scanlon[3]

[1]Center for Space Research, University of Texas at Austin, 3925 West Braker Lane, Ste. 200, Austin, TX 78759, USA

[2]Department of Geological Sciences, Jackson School of Geosciences, University of Texas at Austin, TX 78712, USA

[3]Bureau of Economic Geology, Jackson School of Geosciences, University of Texas at Austin, TX 78712, USA

**Abstract.** The Gravity Recovery and Climate Experiment (GRACE) provides quantitative measures of terrestrial water storage (TWS) change. GRACE data show a significant decrease in TWS in the lower (southern) La Plata river basin of South America over the period 2002 – 2009, consistent with recognized drought conditions in the region. *GRACE data reveal a detailed picture of temporal and spatial evolution of this severe drought event, which suggests that the drought began in lower La Plata in around austral Spring 2008 and then spread to the entire La Plata basin and peaked in austral Fall 2009. During the peak, GRACE data show an average TWS deficit of ~ 12 cm (equivalent water layer thickness) below the 7-year mean, in a broad region in lower La Plata. GRACE measurements are consistent with accumulated precipitation data from satellite remote sensing, and with vegetation index changes derived from Terra satellite observations.* The Global Land Data Assimilation System (GLDAS) model captures the drought event, but underestimates its intensity. *Limited available groundwater level data in* *southern La Plata show significant groundwater depletion, which is likely associated with the drought in this region. GRACE-observed TWS change and precipitation anomalies in the studied region appear to closely correlate with the ENSO climate index, with dry and wet seasons corresponding to La Niña and El Niño events, respectively.*

## 1. Introduction

The La Plata basin is the fifth largest basin in the world and second largest in South America, next to the Amazon basin. With a total area of about 3.5 million km$^2$, covering parts of five countries (Argentina, Uruguay, Paraguay, Brazil, and Bolivia) (Fig. 1), the basin is of great economic and ecological significance, with challenging problems including vulnerability to excess floods and increasing demands as a water resource and source of hydropower [Barros et al., 2004, 2006]. The basin is also home to the Pampas (the dark green area in Fig. 1), one of the world's richest grasslands in terms of size and biodiversity and a major agricultural resource [Viglizzo and Frank, 2006].

The La Plata basin shows evidence of changes that may be identified with long-term climate variation [Barros et al., 2006; Viglizzo and Frank, 2006]. Over the past several decades, the basin has been seeing frequent floods [Minetti et al., 2004; Barros et al., 2006], and more recently has experienced drought. For many areas, especially in the south, the last few years have seen the worst drought in over a century, with official declarations of calamity, a sharp decline in grain and meat output, and economic havoc [Valente, 2009]. The consequences have been especially significant for Argentina, the world's second largest exporter of corn and coarse grains, and the third largest exporter of wheat [FOA, 2000]. The recent drought is likely connected to abnormal climate conditions related to the prolonged 2007/2009 La Niña event [Diaz et al., 1998; de Rojas and Alicia, 2000; Grimm et al., 2000].

Monitoring and quantification of the spatial extent and intensity of drought are limited by conventional data resources (*in situ* meteorological and hydrological observations with sparse spatial and temporal sampling). Deficits in terrestrial water storage (TWS) are particularly difficult to estimate from such data. Drought indices from satellite remote sensing of soil moisture and vegetation change have been used for monitoring drought extent and intensity [e.g., Sims et al., 2002; Wang and Qu, 2007]. Numerical climate and land surface models are valuable in analyzing and diagnosing climate variability, but are imperfect at quantifying extreme climate events, including droughts [Chen et al., 2009].

TWS change is a major component of the global water cycle, and represents the total change of water stored in soil, as snow over land, and in groundwater reservoirs. In a given basin, TWS change reflects the sum of accumulated precipitation, evapotranspiration, and surface and subsurface runoff, and provides a good measure of abnormal climate conditions such as droughts and floods. Quantification of TWS change is difficult because of the lack of fundamental observations of groundwater, soil moisture, snow water equivalent, precipitation, evapotranspiration, and river discharge at basin or smaller scales. Numerical models often poorly estimate TWS changes, especially at interannual and longer time scales [Matsuyama et al., 1995; Chen et al., 2009]. Remote sensing data (e.g., TRMM satellite precipitation data) and *in situ* measurements (e.g., river discharge at gauge stations) are valuable in estimating TWS changes [Crowley et al., 2007; Zeng et al., 2008], but other hydrological parameters are also required (e.g., evapotranspiration).

Satellite gravity measurements from the Gravity Recovery And Climate Experiment (GRACE) provide a means to estimate TWS by direct monitoring of water mass changes. Since March 2002, GRACE measurements of gravity change at monthly intervals [Tapley et al., 2004] have been used to infer mass variation at Earth's surface [Wahr et al., 1998]. GRACE time-variable gravity observations are able to monitor mass changes with a precision of ~ 1.5 cm of equivalent water thickness change [Wahr et al., 2004, 2006]. Early studies applied GRACE data to a variety of problems including TWS change [e.g., Wahr et al., 2004; Tapley et al., 2004; Strassberg et al., 2009, Longuevergne et al., 2010], polar ice sheet balance [e.g., Velicogna and Wahr, 2006; Chen et al., 2006], and oceanic mass change [e.g., Chambers et al., 2004; Lombard et al., 2007].

With improved background geophysical models and data processing techniques [Bettadpur, 2007a, Swenson and Wahr, 2006], reprocessed GRACE release-04 (RL04) gravity fields show significantly improved quality and spatial resolution near 500 km or better [Chen et al., 2008, 2009]. These improvements have enabled applications to a much wider class of problems than during the first few years of the mission, and with nearly eight years of observations, an understanding of interannual and longer-term changes in TWS is now possible. Here we examine TWS change in the La Plata basin using GRACE RL04 data, along with TWS estimates from the global land data assimilation system (GLDAS) [Rodell et al., 2004]. The goal is to quantify the extent and intensity of the recent La Plata basin drought, and to compare GRACE estimates with others from satellite remote sensing and precipitation data and GLDAS.

## 2. Data Processing

### 2.1 TWS Changes from GRACE Gravity Measurements

We use GRACE RL04 time-variable gravity solutions, provided by the Center for Space Research (CSR), University of Texas at Austin [Bettadpur, 2007b]. The 86 approximately monthly gravity solutions cover the period April 2002 through August 2009, and consist of normalized spherical harmonic (SH) coefficients, to degree and order 60. *GRACE SH coefficients are contaminated by noise, including longitudinal stripes (when SH coefficients are converted into mass fields), and other errors, especially at high degrees and orders. The longitudinal stripes have been demonstrated to be associated with unquantified correlations among certain SH coefficients, and removal of these correlations significantly reduces the stripes [Swenson and Wahr, 2006]. For SH orders 6 and above, a least square fit degree 4 polynomial is removed from even and odd degree coefficient pairs [Swenson and Wahr, 2006]. For example, for SH coefficients of order 6 (e.g.,* $C_{n,6}$, $n=6, 7, \ldots, 60$*), we fit a degree 4 polynomial to the even degree pair (e.g.,* $C_{6,6}$, $C_{8,6}$, ..., $C_{60,6}$*) and remove the polynomial fit from the coefficients, and apply the same to the odd degree pair (e.g.,* $C_{7,6}$, $C_{9,6}$, ..., $C_{59,6}$*). We call this decorrelation filter P4M6. After P4M6 filtering, a 300 km Gaussian low-pass filter is applied to further suppress the remaining short-wavelength errors [Jekeli, 1981] and the mean of all 86 monthly solutions is removed from SH coefficient. Monthly mass change fields, expressed as equivalent water layer thickness change on a 1° x 1° grid, are then computed [Wahr et al., 1998].*

*GRACE data have had atmospheric and oceanic mass changes removed using estimates from numerical models during solving GRACE gravity solutions, in a procedure to reduce alias errors in GRACE monthly solutions, due to high frequency atmospheric and oceanic signals [Bettadpur, 2007b]*. Therefore, GRACE mass variations over land should reflect primarily TWS change (including snow/ice) and solid Earth geophysical signals such as postglacial rebound (PGR). Over the La Plata basin, surface mass variations should be dominantly due to near-

surface water storage changes. Errors in GRACE estimates over the La Plata basin are expected to arise from spatial leakage associated with a finite range of SH coefficients, attenuation due to spatial filtering, residual atmospheric signals, and GRACE measurement errors. *Spatial leakage has been a major error source to GRACE estimates, because the truncation of SH coefficients up to degree and order 60 and especially the needed spatial filtering will attenuate the true signal, as a portion of the TWS variance is spread into the surrounding regions (e.g., oceans) (see Fig. 2). Here, we use a 300 km Gaussian low-pass filter (a less strong filter - as 300 km is a relatively shorter spatial scale to GRACE filters) to reduce possible leakage effect, which is likely ~ 5 - 10% of observed signal at seasonal time scales for large basin scale average [Chen et al., 2007].*

## 2.2 TWS Changes from GLDAS Model Estimates

GLDAS ingests satellite- and ground-based observations, using advanced land surface modeling and data assimilation techniques, to generate estimates of land surface states and fluxes [Rodell et al., 2004]. *Precipitation gauge observations, satellite and radar precipitation measurements, and downward radiation flux and analyses from atmospheric data assimilation systems are used as forcing. In the particular simulation used in this study, GLDAS drove the Noah land surface model [Ek et al., 2003], with inputs of precipitation from a spatially and temporally downscaled version of the NOAA Climate Prediction Center's Merged Analysis of Precipitation, and solar radiation data from the Air Force Weather Agency's AGRMET system.* Monthly average soil moisture (2 m column depth) and snow water equivalent were computed from 1979 to present, with TWS at each grid point computed from the sum of soil and snow water. Greenland and Antarctica are excluded because the model omits ice sheet physics. Groundwater is also not modeled by GLDAS.

GLDAS fields need to be spatially filtered in a similar way to the GRACE data for fair comparisons. To accomplish this, GLDAS TWS gridded fields were represented in a SH expansion to degree and order 100, and the P4M6 and 300 km Gaussian smoothing filters were applied. SH coefficients were truncated at degree and order 60, and SH coefficients for degree-0 and degree-1 were set to zero as they are for GRACE fields. Finally, the GLDAS SH expansion was evaluated on a global 1° x 1° grid.

### 2.3 Groundwater Level Data

A collaborative groundwater monitoring project has been set up under the coordination of GEA (Grupo de Estudios Ambientales, Universidad Nacional de San Luis and CONICET) and IyDA-Agritest in the Argentinean Pampas, the southern part of the area of interest. A total of 27 wells (marked by red dots on Figure 2a) are monitoring the shallow groundwater. For each well, monthly water levels were transformed into equivalent water layer using a uniform specific yield (effective porosity) of 0.1 [Aradas et al., 2002]. For each month, water layers were then interpolated using kriging [Wackernagel, 1995] and spatially averaged to extract regional groundwater storage variations.

## 3. Results

### 3.1 GRACE and Climate Model Estimates

At each 1° x 1° grid point there is a time series of TWS variations relative to the mean. We use unweighted least squares to estimate a linear trend, and to evaluate non-seasonal changes, we fit and remove sinusoids at annual, semiannual, and 161-day periods (161 days is the recognized alias period of the S2 tide) [Ray and Luthcke, 2006]. Figure 2a shows mass rates (slope of the linear trend) in the La Plata basin (circled by the gray lines) in units of cm/yr of equivalent water thickness change. GRACE shows significant TWS negative trends (up to ~ 3.5 cm/yr) in the lower La Plata basin during the period April 2002 to August 2009. The area circled by magenta lines identifies the region where negative trends exceed –1 cm/yr. TWS decreases are seen primarily in eastern Argentina and Uruguay. During the same period, the northern La Plata and southern Amazon basins show slight TWS increases. GRACE observations of TWS decrease are consistent with reported drought conditions in the La Plata basin. TWS rate estimates from GLDAS are shown in Figure 2b (the same magenta contour line in Fig. 2a is superimposed here for comparison). GLDAS shows similar TWS decreases in the lower La Plata basin, but the magnitudes are significantly lower than GRACE values (–2.2 vs. –3.5 cm/yr in peak values).

To examine temporal evolution of the drought event, we show in Figure 3a mean TWS changes over the lower La Plata basin (circled by magenta lines in Figs. 2a and 2b) estimated from GRACE and GLDAS. *For a given month, the GRACE uncertainty level is estimated using RMS residuals over the Pacific Ocean in the same latitude zone within the area of 40ºS-25ºS and 180ºE-270ºE. This is an approximation of GRACE uncertainty level, as the true error in GRACE estimates is unknown, due to the lack of other independent measurements of TWS change.* Consistent with the TWS rate maps (Figs. 2a and 2b), both GRACE measurements and GLDAS estimates show a long-term decrease with superimposed seasonal variability. The two estimates (GRACE and GLDAS) agree with each other reasonably well over much of this period. However, GRACE shows much larger TWS increases in the austral spring of 2002 and greater decreases in the falls of 2008 and especially 2009 (the seasons discussed in the present study are referred to the southern hemisphere).

Figure 3b shows non-seasonal GRACE and GLDAS time series, with both indicating a steady decrease in TWS over time. GRACE shows a greater rate of loss. *However, it appears that the large discrepancies in 2002 and 2008/2009 between GRACE and GLDAS primarily drive the slopes difference.* During 2007, both GRACE and GLDAS estimates show significant TWS increase, indicating a reasonably wet season in the lower La Plata. GRACE data indicate that by Fall 2009, average TWS deficit (with respect to the 7 year mean) is about –12 cm, equivalent to ~ 248 Gigatonne (Gt) of water - almost enough water to supply the entire United States for one-half year [Kenny et al., 2009]. The ~ 248 Gt only represents the apparent TWS deficit in the region and the actual amount could be considerably larger, as we have neglected leakage effects from spatial filtering and truncation of spherical harmonic coefficients here, *which are likely not very significant for large regional average as discussed above (see 2.1) [Chen et al., 2007]*.

*The present study appears to reveal a different picture of 'long-term' TWS change in the La Plata basin than that from a previous study [Klees et al., 2008], which doesn't show evident TWS decrease during its studied period (January 2003 – February 2006). The discrepancy is mainly from two factors: 1) the present study uses a much longer record (~ 7 years) of GRACE data than that in Klees et al. (2008) (~ 3 years); and 2) in the present study, we focus in the lower (or southern) La Plata (outlined in magenta in Figs. 2a and 2b), while Klees et al. (2008) targets the entire La Plata basin.*

We compute yearly average GRACE nonseasonal TWS changes for 2003 through 2009 (see Figs. 4a-g). *Each map is the mean over 12 months from July of the previous year to June of current year (solutions for July 2002 and June 2003 are not available, so the 2003 mean is based on 10 solutions). Ocean areas are masked out for clarity. This effectively illustrates the recent* *drought condition in the La Plata basin, which appears to become worsening in Spring 2008, and reach the maximum in Fall 2009 (Fig. 3b). The TWS decrease during Spring 2008 and Fall 2009 is clearly shown by GRACE (Fig. 4f, the 12-month average over July 2008 to June 2009). In 2007 (i.e., average over July 2006 to June 2007), northern La Plata and southern Amazon (and Tocantins Sao Francisco basins) show significant TWS increases, while the lower La Plata TWS remained about average.*

*To further illustrate the temporal and spatial development of this severe drought, we show in Figures 5a - 5l monthly TWS anomalies for a 12-month period from September 2008 to August 2009. Annual and semiannual variations have been removed from each grid point (pixel) using unweighted least squares fit (ocean areas are masked out for clarity). The drought apparently began in lower (southern) La Plata in Spring 2008 (see Figs. 5a-5d), and peaked in Fall 2009 (see Figs. 5h-5j). During the peak months (April – June 2009), the drought spread out to the entire La Plata. By August 2009, the upper (northern) La Plata became mostly normal and even wetter, while the lower La Plata remained in drought condition with decreased magnitude. More recent GRACE data (not shown here) suggest that the drought is completely relieved (and actually wetter than normal) by late 2009.*

### 3.2 Comparisons with Other Observations

From the Global Precipitation Climatology Project (GPCP) daily precipitation estimates (V1.1) [Adler et al., 2003], we compute accumulated yearly (July through June, to match periods represented Figure 4) precipitation totals in the lower La Plata basin (the area circled by magenta lines on Figs. 2a and 2b) for the period 1998 to 2009. Figure 6a shows the result, with GPCP precipitation totals for the GRACE period (after 2002) in gray. Precipitation has decreased since 2003 and average precipitation from 2003 to 2009 is 1072 mm, considerably less than the 1998 to 2002 average of 1265 mm. GRACE observations began during the relatively wet 2003 season, while 2009 recorded the least amount of precipitation during the 12 years, up to 500 mm less than the peak in 2003. These features are qualitatively consistent with GRACE observations (Fig. 3b). There is clear correlation with the La Plata drought indicated in Figure 4. After 2003, the lower La Plata basin experienced mostly dry years, although 2007 is relatively wet. The drought condition strengthened through Fall 2009, consistent with GRACE observations (Fig. 3b and Fig. 4f). *The variation and decrease of yearly precipitations in the lower La Plata basin are supported by similar estimates (see Fig. 6b) of yearly precipitation totals from the Tropical Rainfall Measuring Mission (TRMM) merged monthly precipitation analysis (3B43, V6) [Huffman et al., 2007]. The TRMM 3B43 precipitation estimates are only available since 1998, so the July-through-June yearly total for 1998 (i.e., the yearly total over July 1997 through June 1998) is not available in the TRMM estimates (Fig. 6b). The significantly reduced amount of precipitation in lower La Plata in recent years is clearly the cause of the severe drought condition in the region, and is expected to be associated with decreased evapotranspiration, river discharge, and groundwater recharge.*

Figure 6c shows the NINO3.4 index for 1997 to 2009. NINO3.4 is the average sea surface temperature (SST) anomaly in the tropical Pacific region bounded by 5°N to 5°S, from 170°W to 120°W. This area has large variability on El Niño time scales, and changes in local sea-surface temperature there shift the region of rainfall typically located in the far western Pacific. An El Niño or La Niña event is identified if the 5-month running-average of the NINO3.4 index exceeds +0.4°C for El Niño or –0.4°C for La Niña for at least 6 consecutive months. The NINO3.4 index time series is provided by the Royal Netherlands Meteorological

Institute (http://www.knmi.nl) [Burgers, 1999]. Comparing Figs. 6a and 6c, wet years are well correlated with major El Niño events (e.g., 1997/1998, 2002/2003, and 2007), and dry years with La Niña (e.g., 1999/2000 and 2006, 2008/2009) or weak El Niño events (e.g., 2004) (see Figs. 6a and 6c).

*We also compute yearly precipitation anomaly maps (using GPCP data) in the La Plata river basin and surrounding regions over the same period (July 2002 - June 2009). Following the similar definition used in GRACE yearly TWS maps (Figs. 4a-4g), yearly precipitation totals are the sum from July of the previous year to June of current year. The yearly precipitation anomalies are the yearly totals with respect to the mean of the 7 yearly totals (2003 to 2009), i.e., the average yearly total precipitation over the 7 years is removed from each of the yearly maps (Figs. 7a-7g). Consistent with GRACE observations, during the 2009 season (i.e., July 2008 to June 2009), the La Plata basin, especially the south part, received significantly less amount (up to over 30 cm) of precipitation than the average years, and during the 2003 season (i.e., July 2002 to June 2003), the lower La Plata received up to over 50 cm more precipitation than usual. Both GRACE and precipitation data show that 2007 is a wet season. It's interesting to see that GRACE sees a relatively wet season in lower La Plata in 2004 (Fig. 4b), while precipitation data (Fig. 7b) appear to show an average or even dry season. This may suggest that there may be a lag between precipitation and TWS anomalies. As precipitation is only one of the three major parameters (along with evapotranspiration and runoff) that contribute to TWS change (when groundwater pumping due to human activities is neglected), it is difficult to directly or quantitatively compare GRACE TWS and precipitation anomalies (Fig. 4 vs. Fig.7).*

Figure 8 shows satellite-based normalized difference vegetation index (NDVI) [Sims et al., 2002; Wang and Qu, 2007] for the lower La Plata from the Moderate Resolution Imaging

Spectroradiometer (MODIS) from the NASA Terra satellite. This NDVI map (NASA Earth Observatory) represents the index for January 17–February 1, 2009, relative to the average index during the same period from 2000–2008. The brown color indicates below average vegetation, corresponding to a dry season; white shows normal conditions; and green indicates higher than average, a wet season. Dry conditions are evident in the lower La Plata basin in early 2009. Although NDVI is not a quantitative measure of TWS change, it is useful for monitoring surface drought conditions and is consistent with GRACE estimates.

Currently groundwater level data are not available for the entire area (lower and southern La Plata) examined in this study. Limited groundwater level data are available in a small area in the southern La Plata basin (see Fig. 2a). In this area, the topography is extremely flat and strong interconnections link surface water and the shallow groundwater [Aragon et al., 2010]. Groundwater storage changes were compared with GRACE and GLDAS TWS estimates in the area circled by the red box in Fig. 2a. The discrepancy between GRACE and GLDAS estimates (in this area, Figs. 9a and 9b) appears much greater than that for the broad drought area (shown in Figs. 2 and 3). It is interesting to see that the groundwater storage data from the wells show a significant decreasing trend, consistent with GRACE observations. When groundwater storage data from the wells is added to GLDAS estimates (which does not include a groundwater component), GRACE estimates and the combined GLDAS and well time series show significantly better agreements, at both seasonal and long-term time scales (Fig. 9b). *The decreasing trend in groundwater storage may not necessarily be resulted from the drought condition in the region, and more likely reflects the combined effect from increased groundwater pumping (due to agricultural and industrial usage) and decreased groundwater recharge due to* *the drought condition on the surface. Quantification of these two separate contributions is difficult and also beyond the scope of this study.*

## 4. Conclusions and Discussion

GRACE data indicate a significant decrease in TWS in the lower La Plata basin in recent years and provide a quantitative measure of recent drought conditions. *GRACE TWS estimates reveal  a detailed picture of temporal and spatial evolution of this severe drought event, and suggest that the drought conditions worsened in 2009, with average TWS deficit (with respect to the 7 year mean) reaching in excess of 12 cm equivalent water thickness by Fall 2009 (in a broad region in lower La Plata). GRACE estimates are consistent with GPCP and TRMM precipitation analysis and vegetation index measurements from satellite remote sensing.*

The GLDAS land surface model shows similar TWS changes in the lower La Plata, but with considerably smaller magnitude at longer time scales. The lack of a groundwater component in GLDAS appears to be partly responsible for this discrepancy, at least in the examined area in the south La Plata basin where well water level data are available (Figs. 9a and 9b). Available groundwater data in this region show significant groundwater depletion, *which is likely* associated with the drought. Supplementing GLDAS TWS estimates with groundwater level data significantly improves the agreement with GRACE estimates. Unfortunately, there are no adequate *in situ* TWS measurements to fully validate GRACE estimates. Precipitation data are helpful for qualitatively understanding TWS changes, but cannot be used quantitatively in the absence of evapotranspiration and runoff. This highlights the unique strength of satellite gravity observations in monitoring large spatial scale TWS changes, and providing an independent measurement for calibrating, evaluating, and improving climate and land surface models (Oleson et al., 2008).

*Drought and flood conditions in the La Plata basin appear closely connected to El Niño and La Niña events. These events cause abnormal changes in general circulation patterns and bring increased or decreased precipitation to affected regions [Diaz et al., 1998; de Rojas and Alicia, 2000; Grimm et al., 2000]. This relationship is reinforced by good correlation between precipitation changes in the lower La Plata (Figs. 6a and 6b) and the NINO3.4 SST anomaly index (Fig. 6c) over the period 1997 to 2009. GRACE nonseasonal TWS estimates (Fig. 3b) also correlate well with and the NINO3.4 SST index (Fig. 6c). The 2008/2009 drought in the lower La Plata is likely connected to the 2008/2009 La Niña event. It's interesting to notice that the much stronger 1999/2000 La Niña event also corresponds to a major drought in La Plata [Zanvettor and Ravelo, 2000], however its magnitude (at least in lower La Plata) appears not as significant as the recent drought, as suggested by precipitation data (see Fig. 6a). This indicates that other factors (in addition to 2008/2009 La Niña event) might have contributed to the recent severe drought in lower La Plata as well.*

It is difficult to directly validate GRACE estimates in the absence of adequate *in situ* TWS or related measurements. *Residual variations over the oceans (where the expected signal is zero, if the ocean model estimates used in GRACE dealiasing process are correct) can serve as an approximate of GRACE error [Wahr et al., 2004]. GRACE-observed TWS anomalies in lower La Plata are well over the residuals over the ocean, providing confidence that that the signal is reliable.* The GRACE mission has been extended until at least 2013, and a reprocessed GRACE data set (release 5) will soon incorporate improved background geophysical models and processing methods. These should lead to improved estimates of TWS change for monitoring the climate and providing independent constraints on climate and land surface models.

**Acknowledgments.** We are grateful to "Grupo de Estudios Ambientales, Universidad Nacional de San Luis and CONICET" and IyDA-Agritest for sharing groundwater data. This study was supported by NASA PECASE Award (NNG04G060G), NASA GRACE Program (NNX08AJ84G), and NSF IPY program (ANT-0632195).

**Figure Captions:**

Figure 1. Map of the La Plata basin (outlined in red) in South America.

Figure 2. (a) GRACE mass rates (in cm/yr of water thickness change) in the La Plata basin and surrounding regions from April 2002 to August 2009. A 2-step filtering scheme (P4M6 and 300 km Gaussian smoothing) is applied, as described in the text. The area circled by magenta lines indicates where GRACE rates are in excess of – 1 cm/yr. The red dots mark 27 well locations and the well water level data are used in later analysis. (b) GLDAS average mass rates (in cm/yr of water thickness change) in the same regions and over the same period (April 2002- August 2009). (P4M6 and 300 km Gaussian smoothing are also applied.

Figure 3. a) Comparison of TWS change in the lower La Plata basin (average within the area circled by magenta lines in Figs. 1 and 2) from GRACE (blue curve) and GLDAS (red curve); b) Comparison of nonseasonal TWS changes in the lower La Plata basin from GRACE (blue curve) and GLDAS (red curve). Annual and semiannual signals have been removed using an unweighted least squares fit. The GRACE uncertainty level is estimated using RMS residuals over the Pacific Ocean in the same latitude zone within the area of 40ºS-25ºS and 180ºE-270ºE.

Figure 4. Evolution of yearly TWS deficits (cm of water thickness change) from GRACE in the La Plata river basin and surrounding regions over the 7 years (August 2002 - June 2009). Yearly averages are mean TWS changes from July of the previous year through June (solutions for July 2002 and June 2003 are not available). For example, the 2004 TWS deficit is the mean from July 2003 through June 2004. The mean over the 7 year period is removed from all seven maps. Ocean areas are masked out for clarity.

Figure 5. Monthly TWS anomalies during a 12 months period from September 2008 to August 2009. Annual and semiannual variations have been removed from each grid point (pixel) using unweighted least squares fit. Ocean areas are masked out for clarity.

Figure 6. a) Accumulated yearly (July-June) total precipitation in the lower La Plata basin (circled by the magenta lines in Figs. 2a and 2b) for 1998 – 2009 from GPCP (V1.1). Gray bars are the period spanned by GRACE; b) Accumulated yearly (July-June) total precipitation in the lower La Plata basin (circled by the magenta lines in Figs. 2a and 2b) for 1999 – 2009 from TRMM 3B43 (V6). Gray bars are the period spanned by GRACE; and c) The NINO3.4 index 1997-2009. NINO3.4 is the average sea surface temperature (SST) anomaly in the region bounded by 5°N to 5°S, from 170°W to 120°W. This region has large variability on El Niño time scales, and is associated with the area of rainfall that is typically located in the far western Pacific. The NINO3.4 time series is provided by the Royal Netherlands Meteorological Institute (http://www.knmi.nl).

Figure 7. Evolution of yearly precipitation anomalies (cm of water thickness) from GPCP in the La Plata river basin and surrounding regions over the 7 years (July 2002 - June 2009). Yearly precipitation totals are the sum of July of the previous year through June. The yearly precipitation anomalies are the yearly totals with respect to the mean yearly total over the 7 years

period (i.e., the mean yearly total precipitation over the 7 years is removed from each of the 7 maps).

Figure 8. Normalized difference vegetation index (NDVI) for the lower La Plata basin from the Moderate Resolution Imaging Spectroradiometer (MODIS) and NASA Terra satellite observations. This shows the index for January 17–February 1, 2009, relative to the average index of 2000- 2008. Brown indicates vegetation below average levels, associated with a dry season; white shows normal conditions; and greens shows a higher than average index, associated with a wet season. (NASA image from http://earthobservatory.nasa.gov/NaturalHazards/view.php?id=37239).

Figure 9. (a) Comparison of TWS changes from GRACE, GLDAS, and average groundwater storage change from 27 wells (marked by red dots in Fig. 2a) in the south La Plata basin. GRACE and GLDAS time series are the average estimates within the area circled by red box in Fig. 2a. (b) Similar as (a), but with seasonal variations removed using unweighted least squares fit. A specific yield (effective porosity) of 10% is applied when computing groundwater storage from well water level data.

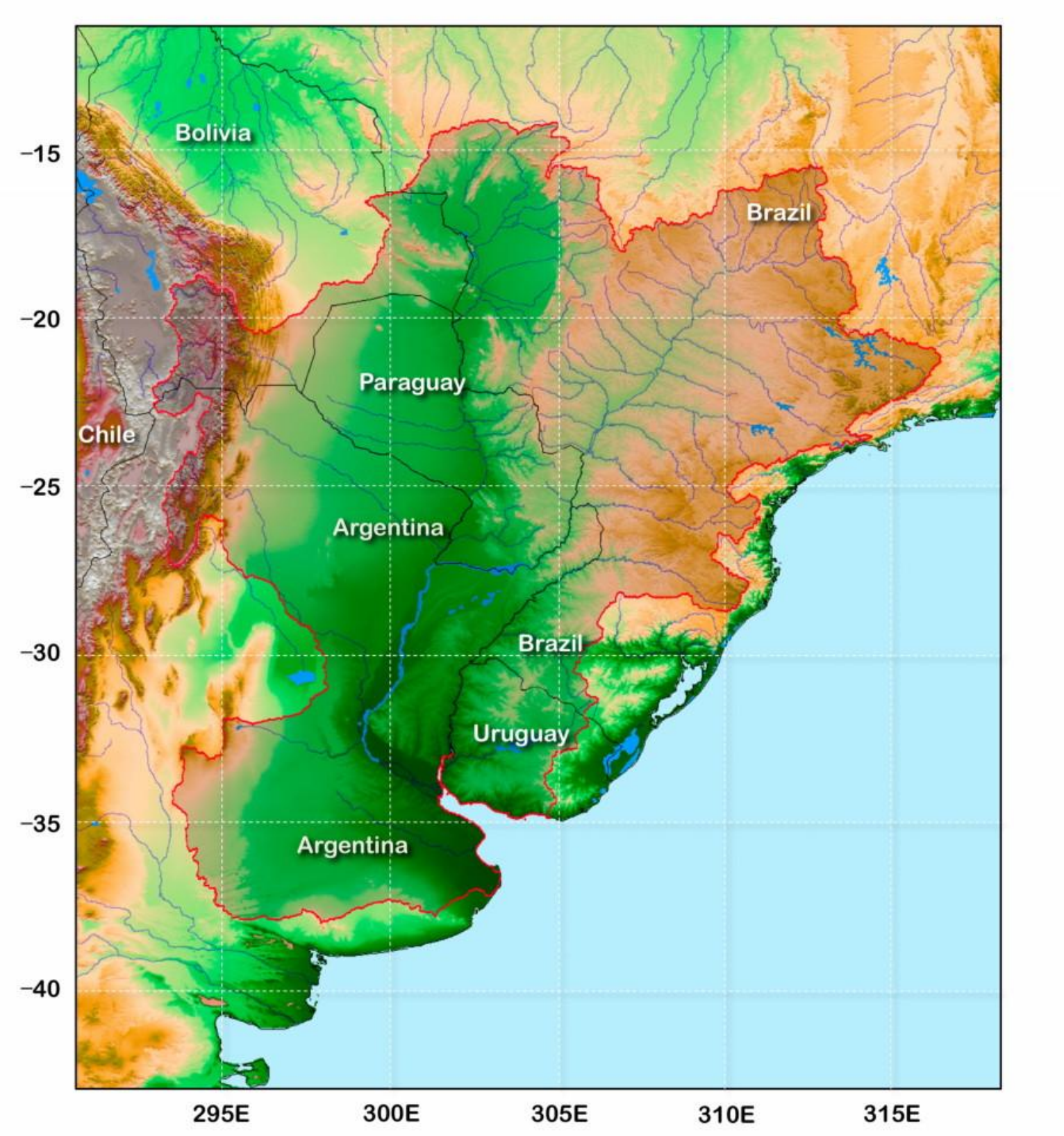

Figure 1

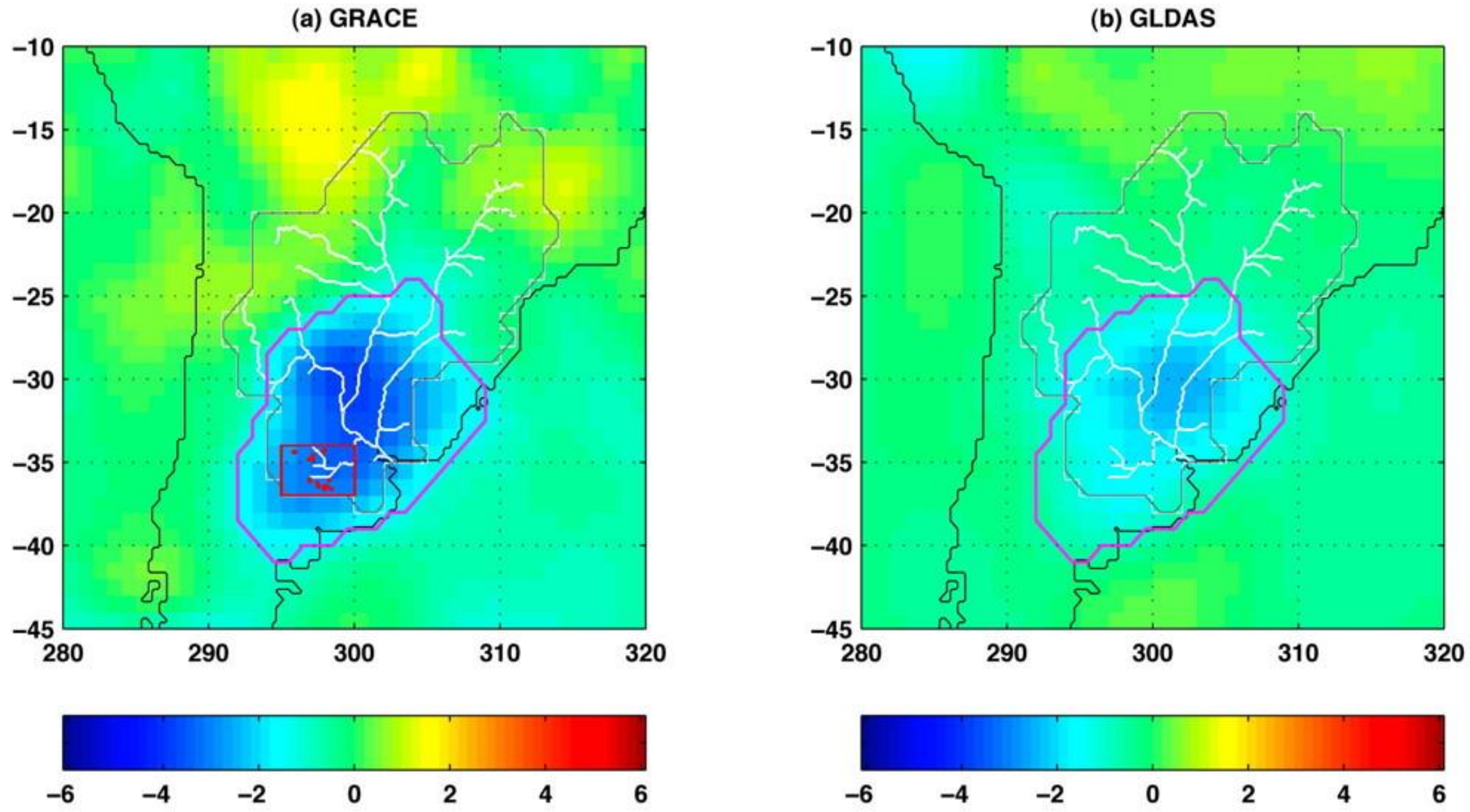

Figure 2

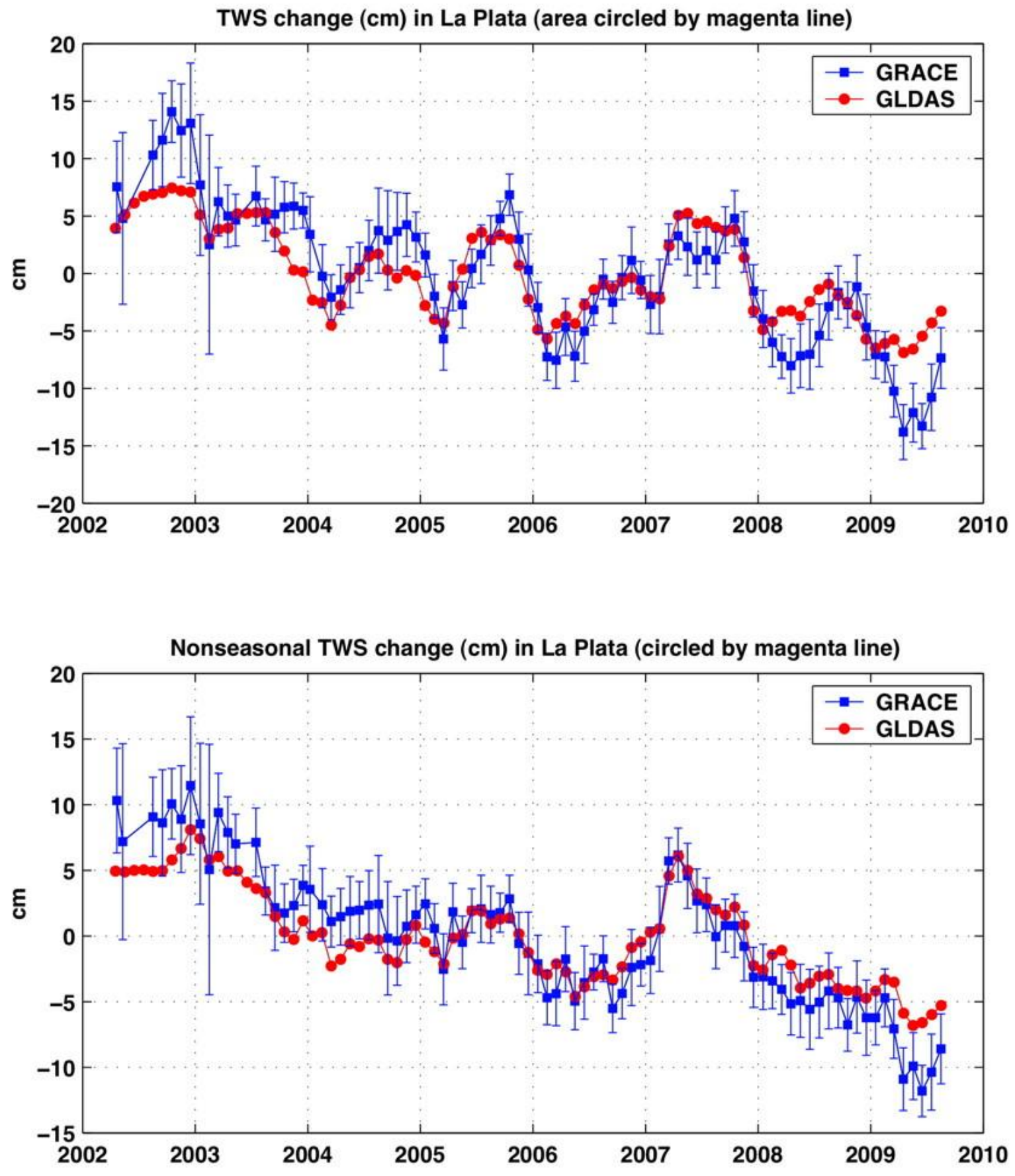
TWS change (cm) in La Plata (area circled by magenta line)
GRACE
GLDAS
cm
Nonseasonal TWS change (cm) in La Plata (circled by magenta line)
GRACE
GLDAS
cm

Figure 3

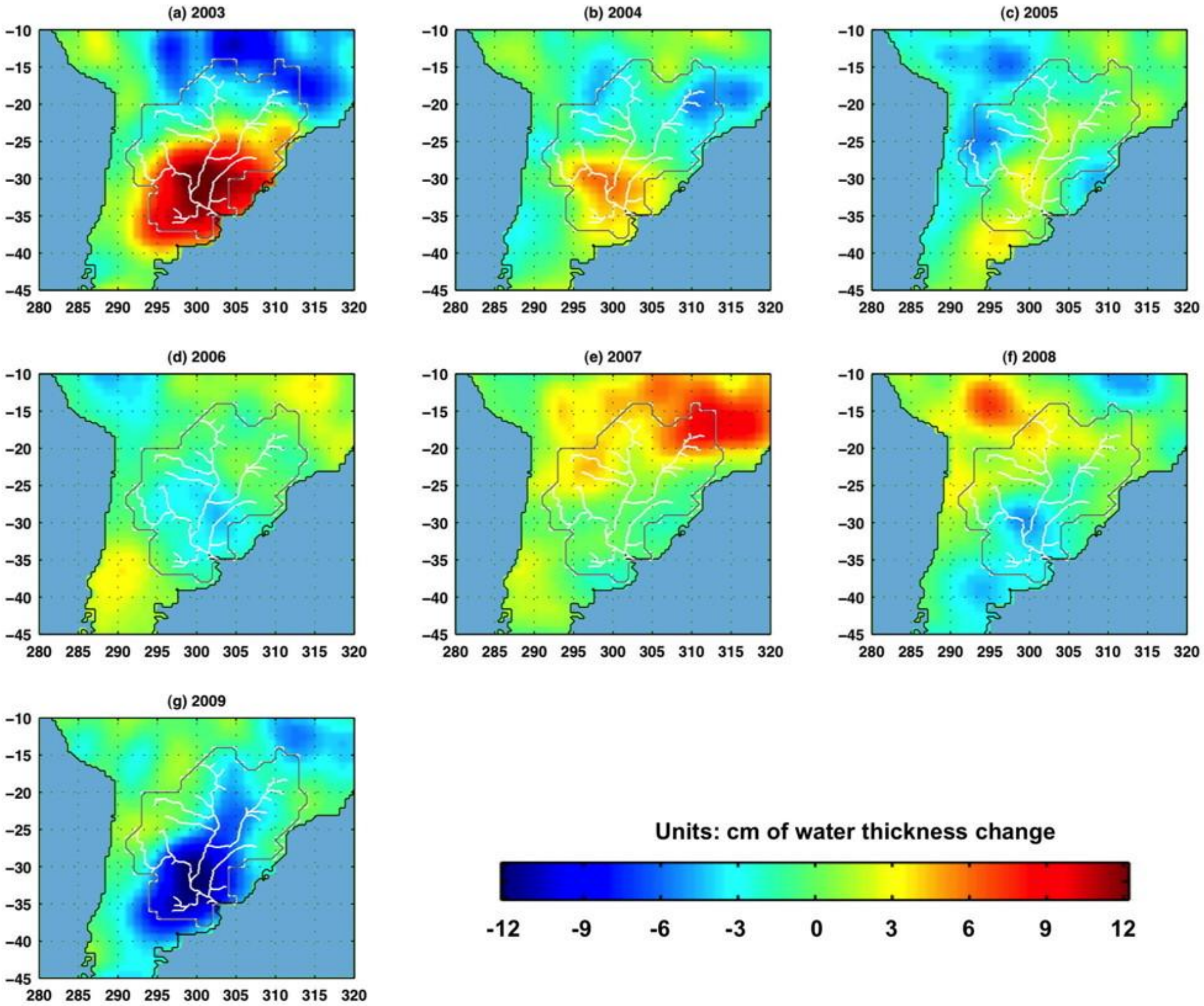
(a) 2003
(b) 2004
(c) 2005
(d) 2006
(e) 2007
(f) 2008
(g) 2009
-10
-15
-20
-25
-30
-35
-40
-45
280 285 290 295 300 305 310 315 320
Units: cm of water thickness change
-12 -9 -6 -3 0 3 6 9 12

Figure 4

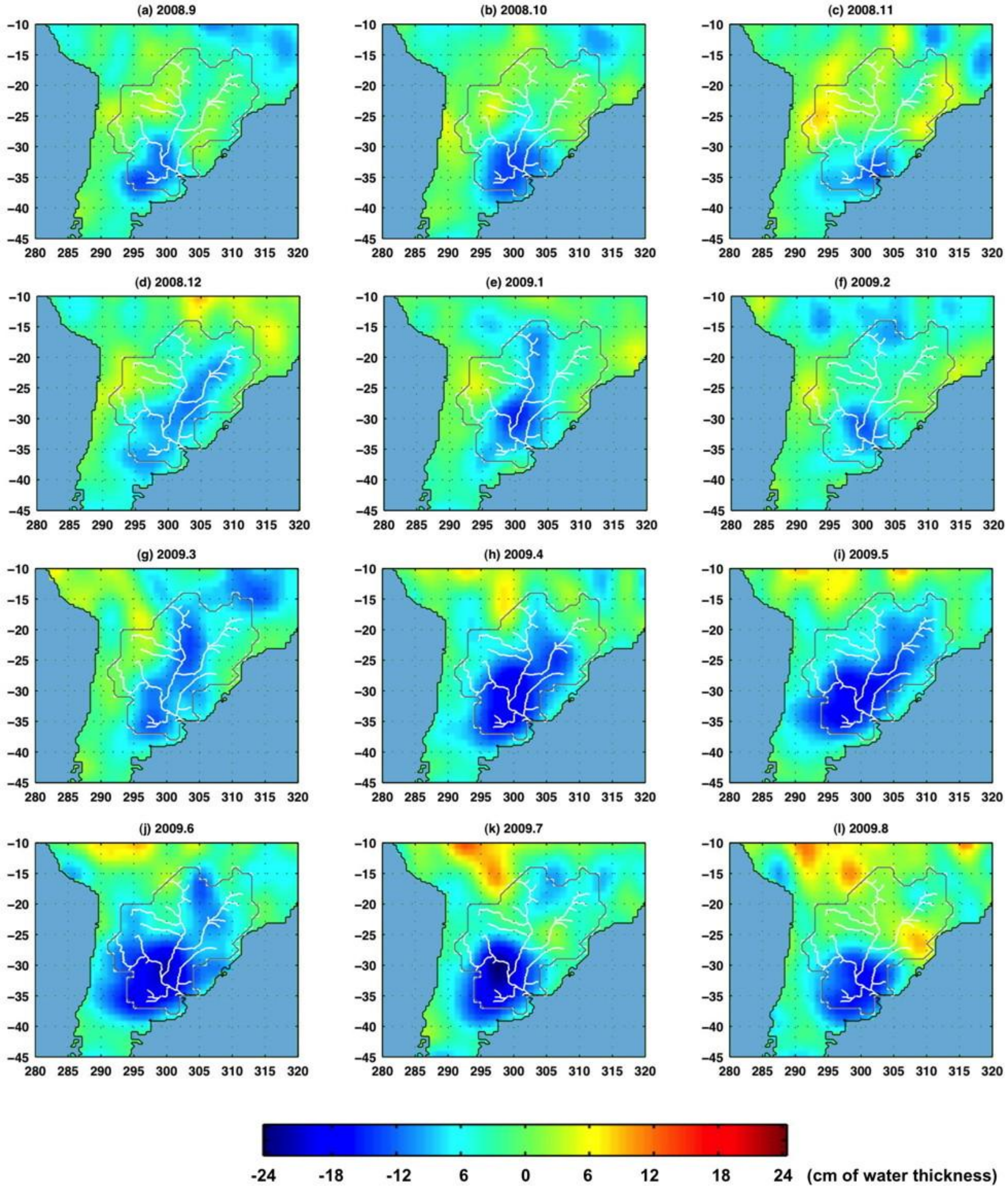
(a) 2008.9
(b) 2008.10
(c) 2008.11
(d) 2008.12
(e) 2009.1
(f) 2009.2
(g) 2009.3
(h) 2009.4
(i) 2009.5
(j) 2009.6
(k) 2009.7
(l) 2009.8
-10
-15
-20
-25
-30
-35
-40
-45
280 285 290 295 300 305 310 315 320
-24 -18 -12 6 0 6 12 18 24 (cm of water thickness)

Figure 5

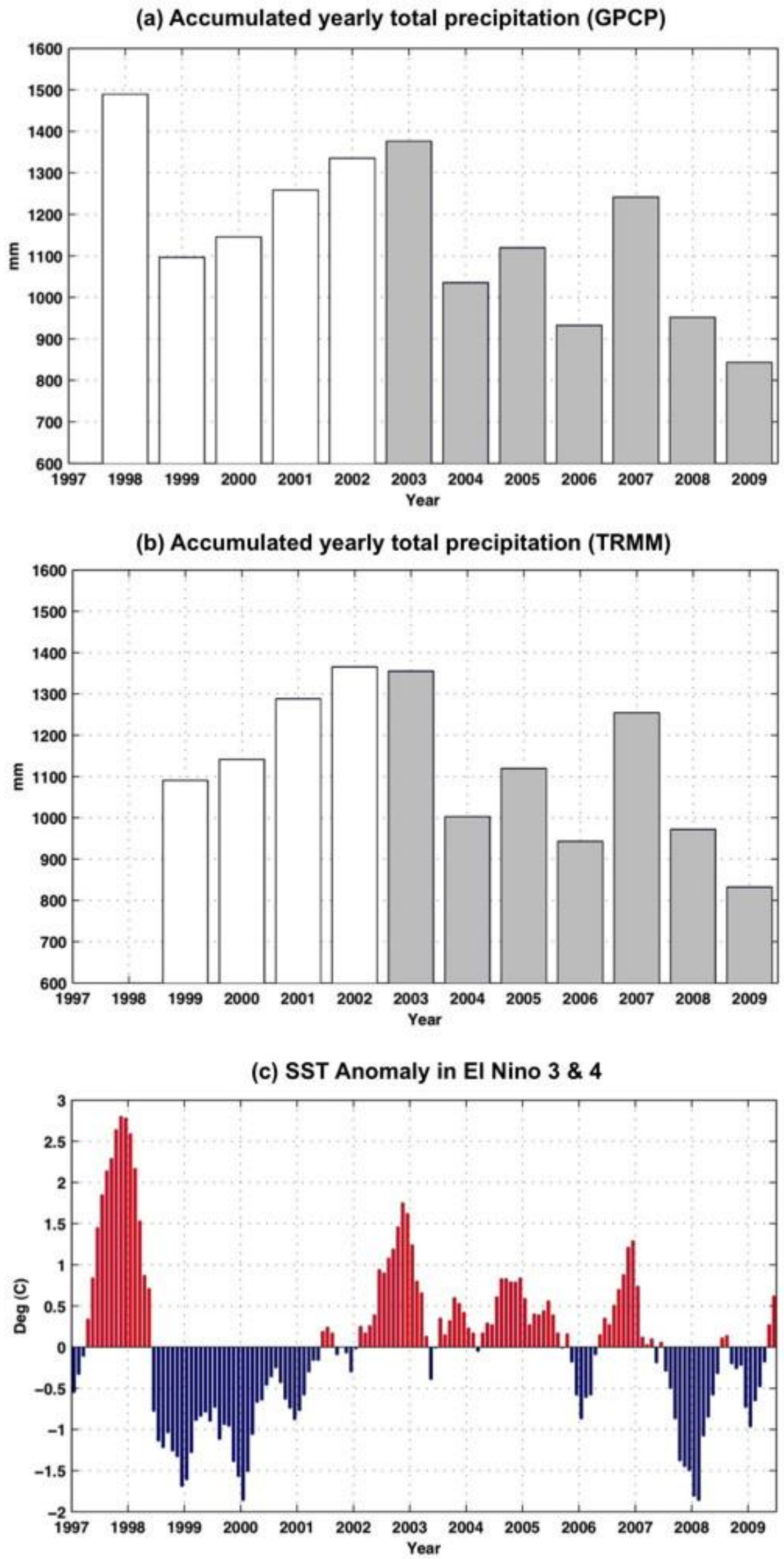
(a) Accumulated yearly total precipitation (GPCP)
mm
1600
1500
1400
1300
1200
1100
1000
900
800
700
600
1997 1998 1999 2000 2001 2002 2003 2004 2005 2006 2007 2008 2009
Year
(b) Accumulated yearly total precipitation (TRMM)
mm
1600
1500
1400
1300
1200
1100
1000
900
800
700
600
1997 1998 1999 2000 2001 2002 2003 2004 2005 2006 2007 2008 2009
Year
(c) SST Anomaly in El Nino 3 & 4
Deg (C)
3
2.5
2
1.5
1
0.5
0
-0.5
-1
-1.5
-2
1997 1998 1999 2000 2001 2002 2003 2004 2005 2006 2007 2008 2009
Year

Figure 6

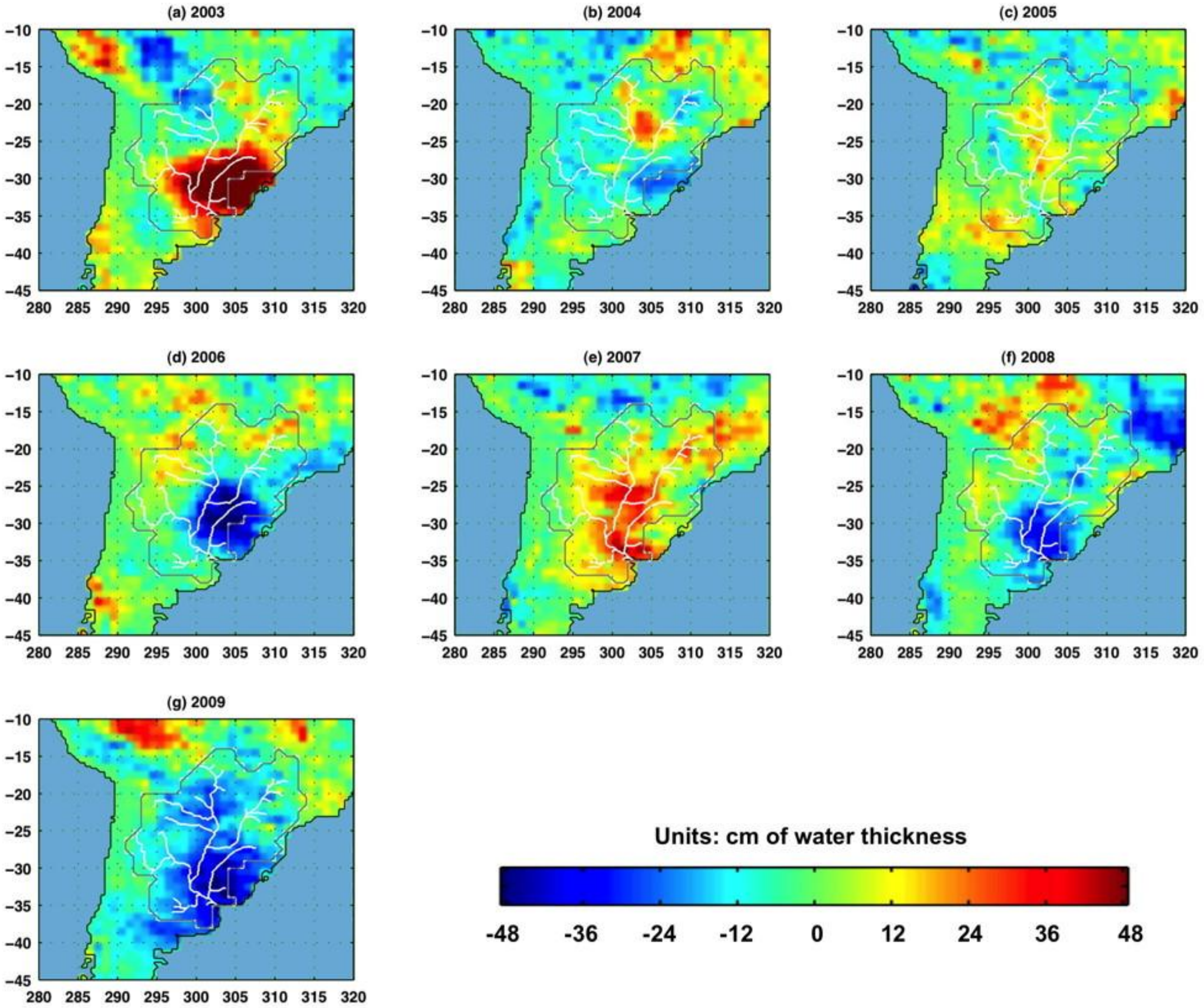
(a) 2003
(b) 2004
(c) 2005
(d) 2006
(e) 2007
(f) 2008
(g) 2009
-10
-15
-20
-25
-30
-35
-40
-45
280 285 290 295 300 305 310 315 320
Units: cm of water thickness
-48 -36 -24 -12 0 12 24 36 48

Figure 7

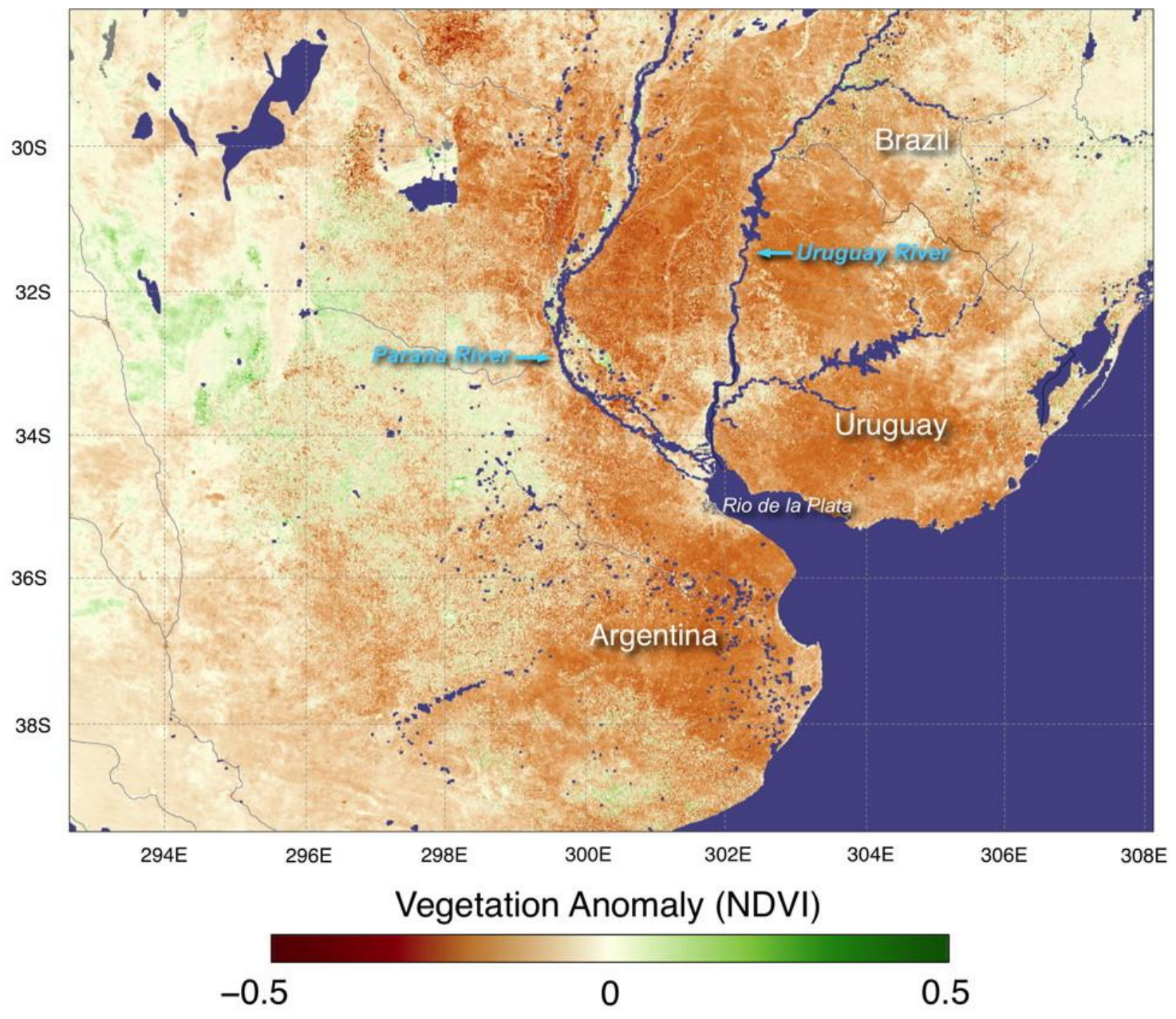

Figure 8

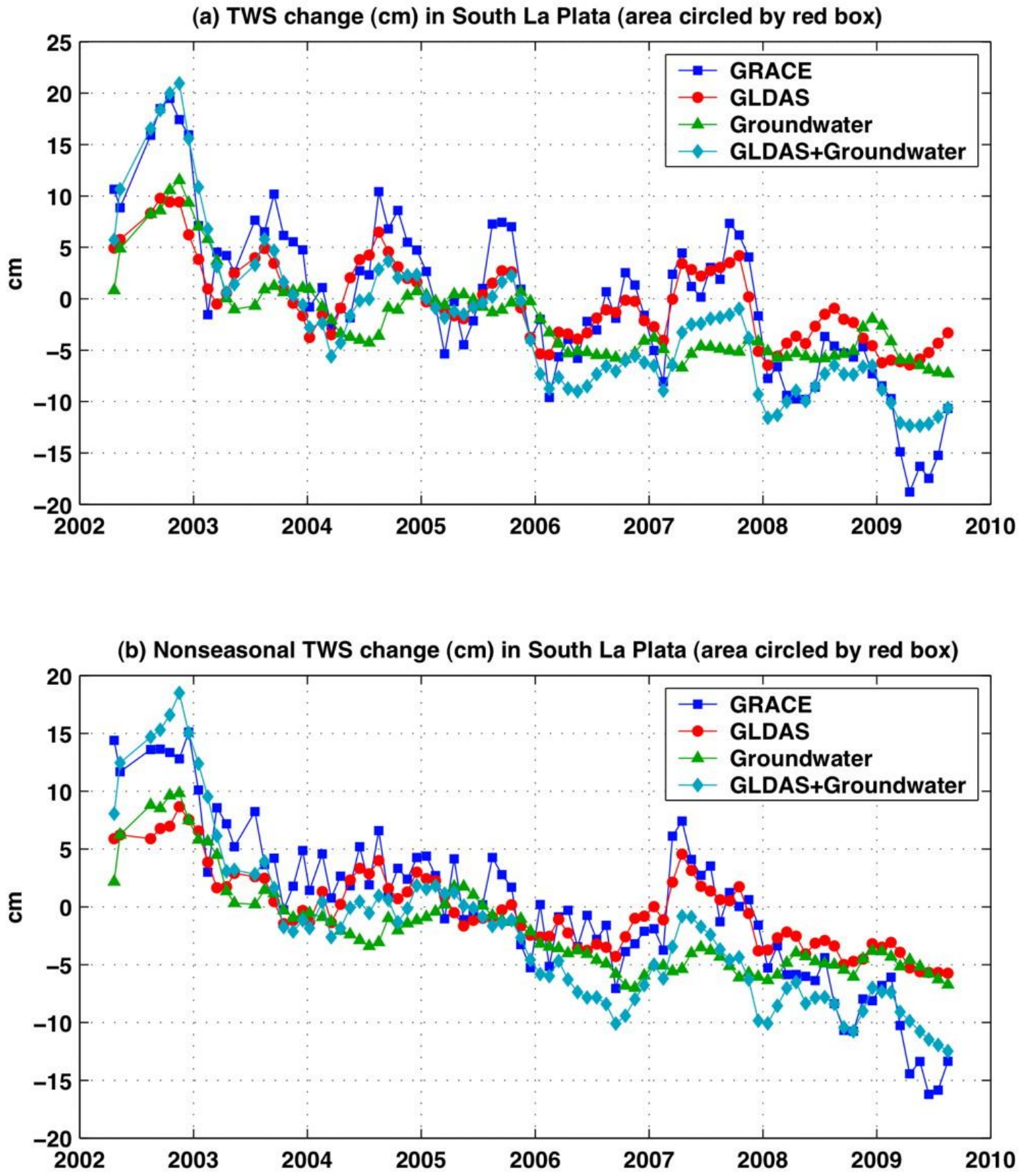
(a) TWS change (cm) in South La Plata (area circled by red box)
GRACE
GLDAS
Groundwater
GLDAS+Groundwater
cm
25
20
15
10
5
0
−5
−10
−15
−20
2002
2003
2004
2005
2006
2007
2008
2009
2010
(b) Nonseasonal TWS change (cm) in South La Plata (area circled by red box)
GRACE
GLDAS
Groundwater
GLDAS+Groundwater
cm
20
15
10
5
0
−5
−10
−15
−20
2002
2003
2004
2005
2006
2007
2008
2009
2010

Figure 9